\documentclass[aps,12pt]{revtex4} 
\usepackage{graphicx}				
\usepackage{amssymb}

\begin{document}

\title{From bricks to quasinormal modes:  A new perspective on black hole entropy.}

\author{Michele Arzano} 
\email{michele.arzano@roma1.infn.it}
\affiliation{Dipartimento di Fisica and INFN,\\ ``Sapienza" University of Rome,\\ P.le A. Moro 2, 00185 Roma, EU }

\author{Stefano Bianco}
\email{stefanobianco3@gmail.com}
\affiliation{Dipartimento di Fisica and INFN,\\ ``Sapienza" University of Rome,\\ P.le A. Moro 2, 00185 Roma, EU }

\author{Olaf Dreyer}
\email{olaf.dreyer@gmail.com}
\affiliation{Dipartimento di Fisica and INFN,\\ ``Sapienza" University of Rome,\\ P.le A. Moro 2, 00185 Roma, EU }

\begin{abstract}

Calculations of black hole entropy based on the counting of modes of a quantum field propagating in a Schwarzschild background need to be regularized in the vicinity of the horizon.  To obtain the Bekenstein-Hawking result the short distance cut-off needs to be fixed by hand.  In this note we give an argument for obtaining this cut-off in a natural fashion.  We do this by modelling the black hole by its set of quasinormal modes.  The horizon then becomes a extended region:  \emph{the quantum ergosphere}.  The interaction of the quantum ergosphere and the quantum field provides a natural regularization mechanism.  The width of the quantum ergosphere provides the right cut-off for the entropy calculation.  We arrive at a dual picture of black hole entropy.  The entropy of the black hole is given both by the entropy of the quantum field in the bulk and the dynamical degrees of freedom on the horizon.  
\begin{center}
{\it Honorable Mention in the 2013 Gravity Research Foundation Essay Competition.}\\
\end{center}
\end{abstract}

\maketitle

What is the nature of black hole entropy?  What kind of micro-states are being counted in the Bekenstein-Hawking entropy
\begin{equation}
S_{\text{BH}}=\frac{A}{4l^2_{\text{Planck}}}\ ?
\end{equation}
There are two schools of thought.  One school holds that black hole entropy affords us a rare look into quantum gravity.  In this view $S_{\text{BH}}$ counts the quantum mechanical microstates that all represent the same classical black hole.  This counting has been performed in different candidate theories of quantum gravity \cite{strings1, lqg}.  In the alternative view the entropy that is calculated has nothing to do with the quantum geometry of the black hole but is instead the entropy of the quantum field in the black hole background.  This point of view was first proposed by 't Hooft \cite{tHooft:1985vk}.  Independently Sorkin et al. \cite{Bombelli:1986rw} suggested that the entanglement entropy of the quantum field is responsible for the Bekenstein-Hawking result.  Further results showed that the entanglement entropy is naturally proportional to the area separating the entangled regions \cite{Srednicki:1993dl}.  It can be shown that 't Hooft's calculation and the entanglement calculation actually coincide, see e.g. \cite{Brustein:2011te}.  

One problem with the calculation of the entropy via the entropy of the quantum field on the black hole spacetime is that the result usually diverges and requires regularization.  In 't Hooft calculation the entropy is given by \footnote{We consider only one particle species here.  The entropy is otherwise proportional to the number of particle species.}
\begin{equation}\label{eqn:entropy}
S = \frac{8\pi^3}{45 h} 2 M \left( \frac{2M}{\beta}\right)^3.
\end{equation}
Here $\beta$ is the inverse Hawking temperature.  Of great importance for what follows is the small distance $h$.  It is clear from equation (\ref{eqn:entropy}) that the entropy diverges if $h$ goes to zero.  The distance $h$ thus regularizes the entropy calculation.  It arises in the calculation of the number $n$ of modes with an energy less than $E$.  Given this energy $E$ and the angular momentum number $l$ the wave number of the mode is given by
\begin{equation}
k^2 = \frac{1}{(1 - \frac{2M}{r})} \left(\left(1 - \frac{2 M}{r}\right)^{-1} E^2 - \frac{l (l+1)}{r^2} - m^2\right).
\end{equation}
The number $n$ can then be calculated from the total phase divided by $\pi$:
\begin{equation}\label{eqn:n}
n = \frac{1}{\pi}\int_{2M + h}^L dr\ k(r,l,E)
\end{equation}
The function $k^2$ has a singularity at $r=2M$ which causes the integral (\ref{eqn:n}) to diverge in the limit where $h$ goes to zero.  The key point noted by 't Hooft is that if $h$ is fixed to the value
\begin{equation}
\frac{1}{720 \pi M}
\end{equation}
then equation (\ref{eqn:entropy}) gives the correct Bekenstein-Hawking entropy.  In this approach the length scale $h$ has to be fixed by hand to get the desired result.  It is important to note here that $h$ not only has to be small but also has to scale in a non-trivial way with the mass $M$ of the black hole.

In this essay we will argue that the interaction of the quantum field with the quantum black hole provides an independent way to fix $h$.  We will model the quantum back-reacting black hole horizon via its set of quasinormal modes (QNMs), an idea which dates back to York \cite{YorkJr:1983vu}.  The black hole horizon then seizes to be a sharply defined three-manifold and instead becomes an extended region.  The thickness of this region is proportional to $M^{-1}$ and it is this thickness that we will chose for the parameter $h$.  This then gives us the correct entropy.
\\

\noindent
{\it What happens to a black hole horizon when it radiates?}\\
The event horizon is composed of those light rays that are just able to escape to null infinity.  In the case of the static Schwarzschild black hole the event horizon is foliated by apparent horizons.  The radii of apparent horizons and event horizons are thus equal.  This changes when the black hole absorbs matter from outside the horizon.  Imagine a light ray that is close to but outside the horizon.  If nothing happens to the black hole it will escape to infinity.  If,  on the other hand,  matter is dropped into the black hole the light ray might no longer be able to escape.  It follows that the event horizon is \emph{larger} than the apparent horizon if matter falls into the horizon in the \emph{future}.  Here we are interested in the reverse process:  the black hole is loosing mass.  Because of Hawking radiation it has the luminosity \footnote{This is the Stefan-Boltzman luminosity for a black body with the Hawking temperature $T_{\text{H}}$ and area $A = 4\pi(2M)^2$. } 
\begin{equation}\label{eqn:hawkingl}
L = \frac{\hbar B}{M^2},
\end{equation}
where we have combined all the numerical factors into the constant $B$.  The black hole is loosing mass and because of that the event horizon is now \emph{inside} the apparent horizon.  York \cite{YorkJr:1983vu} estimated that the radius of the event horizon of a black hole that is radiating with a luminosity $L$ is given by
\begin{equation}
r_{\text{EH}} = 2 M (1 - 4 L).
\end{equation}
He called the region between the event horizon and the apparent horizon the \emph{quantum ergosphere}.  The thickness of the quantum ergosphere is given by
\begin{equation}
\Delta r_{\text{qe}} = 8 M L\, .
\end{equation}
This relation is valid for any kind of luminosity small enough that $\mathcal{O}(L^2)$ effects can be neglected.  If we now substitute the Hawking value for the luminosity from equation (\ref{eqn:hawkingl}) we obtain
\begin{equation}
\Delta r_{\text{qe}} = \frac{8 \hbar}{M}.
\end{equation}
Remarkably this is the correct $M$ dependence for the small distance cut-off $h$ near the horizon.\\ 

\noindent
{\it A wall of quasi-normal modes}\\
The observation above is very suggestive.  We will now make a more concrete model of the quantum black hole.  In the spirit of \cite{YorkJr:1983vu} we will model the quantum black hole by a \emph{collection of damped harmonic oscillators with the quasinormal mode frequencies of the black hole.}

\begin{figure}[th]
\begin{center}
\includegraphics[width=12cm]{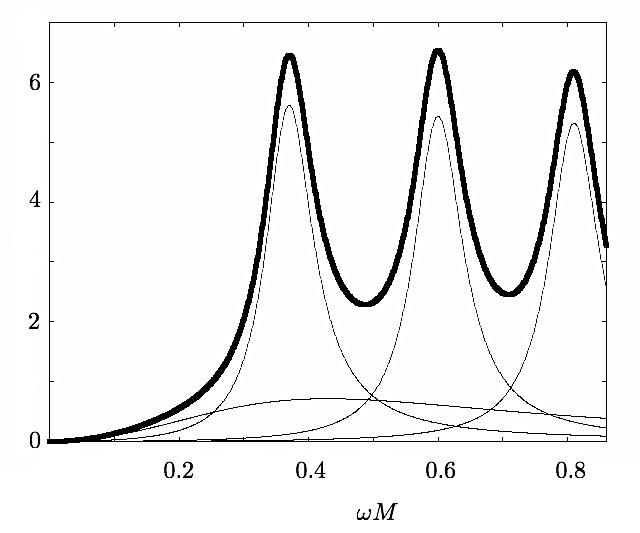}
\caption{The resonant frequencies of the QNMs overlap sufficiently to form an impenetrable wall.  Here we show the absorption of energy for the QNMs with $l=2,3,4$ and $n=0$.  This wall replaces the brick wall of 't Hooft.}\label{fig:qnmwall}
\end{center}
\end{figure}

The quasinormal modes of the Schwarzschild black hole are labeled by a pair $(n,l)$ of positive integers,  with $l\ge 2$ and $n\ge 0$.  The least damped modes are the $n=0$ modes.  We will focus on them for now.  The large $l$ limit of the frequencies $\omega$ for these modes are given by \cite{Kokkotas:1999uk}
\begin{equation}
3\sqrt{3} M \omega = \left(l + \frac{1}{2}\right) - \frac{i}{2}.
\end{equation}
From this we can read off the $Q$-factor for the individual modes:
\begin{equation}
Q = l +1
\end{equation}
The corresponding width of the resonance is then
\begin{eqnarray}
\Delta \omega & = & \omega_R/Q \\
	& = & \frac{1}{3\sqrt{3} M}.
\end{eqnarray}
We see that the resonances of the individual QNMs overlap each other (see figure \ref{fig:qnmwall}).  Just looking at this very restricted set of QNMs we see that for all frequencies above $0.37 M^{-1}$ there is thus a QNM that will be excited by this frequency.  The quasinormal modes form a wall of resonant frequencies.  This wall replaces the brick wall of 't Hooft.
\\

\noindent
{\it The quantum ergosphere as a brick wall}\\
Because of the presence of the quasinormal modes the quantum field will not be able to oscillate freely in the vicinity of the black hole horizon.  We will then regularize the divergent expression (\ref{eqn:n}) by cutting off the integral at the Schwarzschild radius $2M$ plus the average amplitude of the horizon oscillations due to the quasinormal modes.  

Since $A = 4\pi r^2$ we have $\Delta A = 8\pi r \Delta r.$  In \cite{YorkJr:1983vu} York argues that the average area of the event horizon is
\begin{equation}
\bar A = 4\pi (2M)^2 \left( 1 - \frac{1}{2}\epsilon^2 (1 + 16(M\omega)^2)^{-1}\right),
\end{equation}
This gives the radius change
\begin{equation}
\Delta r = \frac{1}{2}M\epsilon^2 (1 + 16(M\omega)^2)^{-1}.
\end{equation}
If we just focus on the $l=2$ mode we have $M\omega = 0.37$ and $\epsilon^2$ is given by\cite{YorkJr:1983vu}
\begin{equation}
\epsilon^2 = 2 \times 10^{-4} \frac{\hbar}{M^2}.
\end{equation}
This gives for $\Delta r$:
\begin{equation}
\Delta r = 0.3 \times 10^{-4} \frac{\hbar}{M}
\end{equation}
Which is the same order of magnitude as the result for $h$ quoted above.\\

\noindent
{\it Black hole entropy: inside or out?}\\
In this essay we make the simple but striking observation that the back-reaction of quantum radiance on a black hole horizon naturally regularizes the calculation of black hole entropy from quantum field theory.  \emph{The result reproduces the right area dependence of the Bekenstein-Hawking relation}.\\
We modelled the geometry of the fluctuating horizon in terms of its quasinormal modes which form a {\it quantum ergosphere}.  We argue that the presence of the quantum ergosphere suppresses the oscillations of the quantum field in the vicinity of the black hole horizon.  This suppression renders 't Hooft's entropy calculation finite.\\
The view of black hole entropy that emerges here is a synthesis of the two prevalent views.  Instead of just having either the degrees of freedom of the quantum field or the degrees of freedom of the quantum geometry we have both.  The remarkable feature of our result is that there is a non-trivial interplay between in and out degrees of freedom.  To have a finite result for one the other must be present.

\end{document}